\documentclass{jpsj-suppl}
\usepackage{txfonts} %Please comment out this line unless the txfonts package is availabe in your LaTeX system.

\title{Pinning Down the Superfluid and Nuclear Equation of State and Measuring
Neutron Star Mass Using Pulsar Glitches}

\author{Wynn C.~G. \textsc{Ho}$^{1}$,
 Crist\'{o}bal M. \textsc{Espinoza}$^{2}$, Danai \textsc{Antonopoulou}$^{1}$,
 and Nils \textsc{Andersson}$^{1}$}

\inst{$^{1}$Mathematical Sciences, Physics \& Astronomy, and STAG Research
Centre, University of Southampton, Southampton, SO17 1BJ, United Kingdom \\
$^{2}$Departamento de F\'{i}sica, Universidad de Santiago de Chile, Avenida
Ecuador 3493, Estaci\'{o}n Central, Santiago, Chile}

\email{wynnho@slac.stanford.edu}

\recdate{July 18, 2016}

\abst{
Pulsars are rotating neutron stars that are renowned for their timing precision,
although glitches can interrupt the regular timing behavior when these stars
are young. Glitches are thought to be caused by interactions between normal and
superfluid matter in the star.  We update our recent work on a new technique
using pulsar glitch data to constrain superfluid and nuclear equation of state
models, demonstrating how current and future astronomy telescopes can probe
fundamental physics such as superfluidity near nuclear saturation and matter
at supranuclear densities.
Unlike traditional methods of measuring a star's mass by its gravitational
effect on another object, our technique relies on nuclear physics knowledge and
therefore allows measurement of the mass of pulsars which are in isolation.}

\kword{equations of state of neutron-star matter, neutron stars, pulsars,
quantum fluids}

\begin{document}
\maketitle

\vspace*{-1.0cm}
\section{Pulsar Spin-down and Superfluid Model of Glitch Spin-up}

Rotating neutron stars, or pulsars, are born in the collapse and supernova
explosion at the end of a massive star's life.
More massive than the Sun but only $\sim\!20$~km in diameter, pulsars are
primarily composed of neutron-rich matter near and above nuclear densities.
Pulsars emit beamed electromagnetic radiation, and
this loss of energy comes at the expense of the pulsar's rotational energy,
causing the star to spin more slowly over time.
While most isolated pulsars are observed to rotate very stably with a spin
period between $\sim\!1$~ms and $10$~s, the timing behavior of many young
pulsars is interrupted by sudden changes, so-called glitches.
During glitches, the pulsar spin rate suddenly increases over a very short
time and then usually relaxes to its pre-glitch rate over a longer time.

Glitches are thought to be the manifestation of a neutron superfluid in the
pulsar's inner crust,
which contains neutron-rich nuclei embedded in a sea of free neutrons.
These free neutrons are expected to be in a superfluid state because the
critical temperature $T_{\rm c}$ (below which neutrons become superfluid)
is well above the typical crust temperature of pulsars.
Unlike normal matter, superfluid matter
in the inner crust rotates by forming vortices whose areal density determines
the spin rate of the superfluid.
To decrease its spin rate, superfluid vortices must move
so that the areal density decreases.
In the inner crust of a neutron star, these vortices are usually pinned
to the nuclei of normal matter \cite{andersonitoh75}.
While the rest of the star slows down owing to electromagnetic energy loss,
the neutron superfluid does not.
As a result, this superfluid can act as a reservoir of angular momentum.
Over many pulsar rotations,
an increasing lag develops between the stellar spin rate and that
of the neutron superfluid in the inner crust.
When this lag exceeds a critical value, superfluid vortices 
unpin and transfer their angular momentum to the rest of the star, causing
the stellar rotation rate to increase and producing what we observe as a
glitch \cite{baymetal69,andersonitoh75}.
Thus glitches provide a valuable measure of the amount of superfluid
angular momentum, or equivalently moment of inertia, in neutron stars.
To probe superfluid and nuclear equation of state (EOS) properties with
pulsar glitches, three questions need answering.

\section{Observational and Theoretical Constraints on Pulsar Moment of Inertia}
\subsection{Question 1: How much angular momentum or moment of inertia do
 glitches require?}

For glitching pulsars, $G\equiv 2\tau_{\rm c}\langle A\rangle$ is the
measured parameter that is to be compared to theoretical models.
Here $\tau_{\rm c}=\Omega/2\dot{\Omega}$,
$\Omega$ and $\dot{\Omega}$ are the pulsar spin rate and its time derivative,
respectively, and $\langle A\rangle=(1/t_{\rm obs})\sum\Delta\Omega/\Omega$
is the average activity parameter
and the summation is over each glitch with spin rate change $\Delta\Omega$
during an observation time span $t_{\rm obs}$.
$G=1.62\pm0.03\%$ for the Vela pulsar, and $G=0.875\pm0.005\%$ for
PSR~J0537$-$6910.
The fractional moment of inertia of the superfluid must exceed $G$ in order
for there to be sufficient angular momentum to be transferred during glitches.

\begin{figure}[tbh]
\vspace*{-0.6cm}
\begin{center}
\includegraphics[scale=0.30]{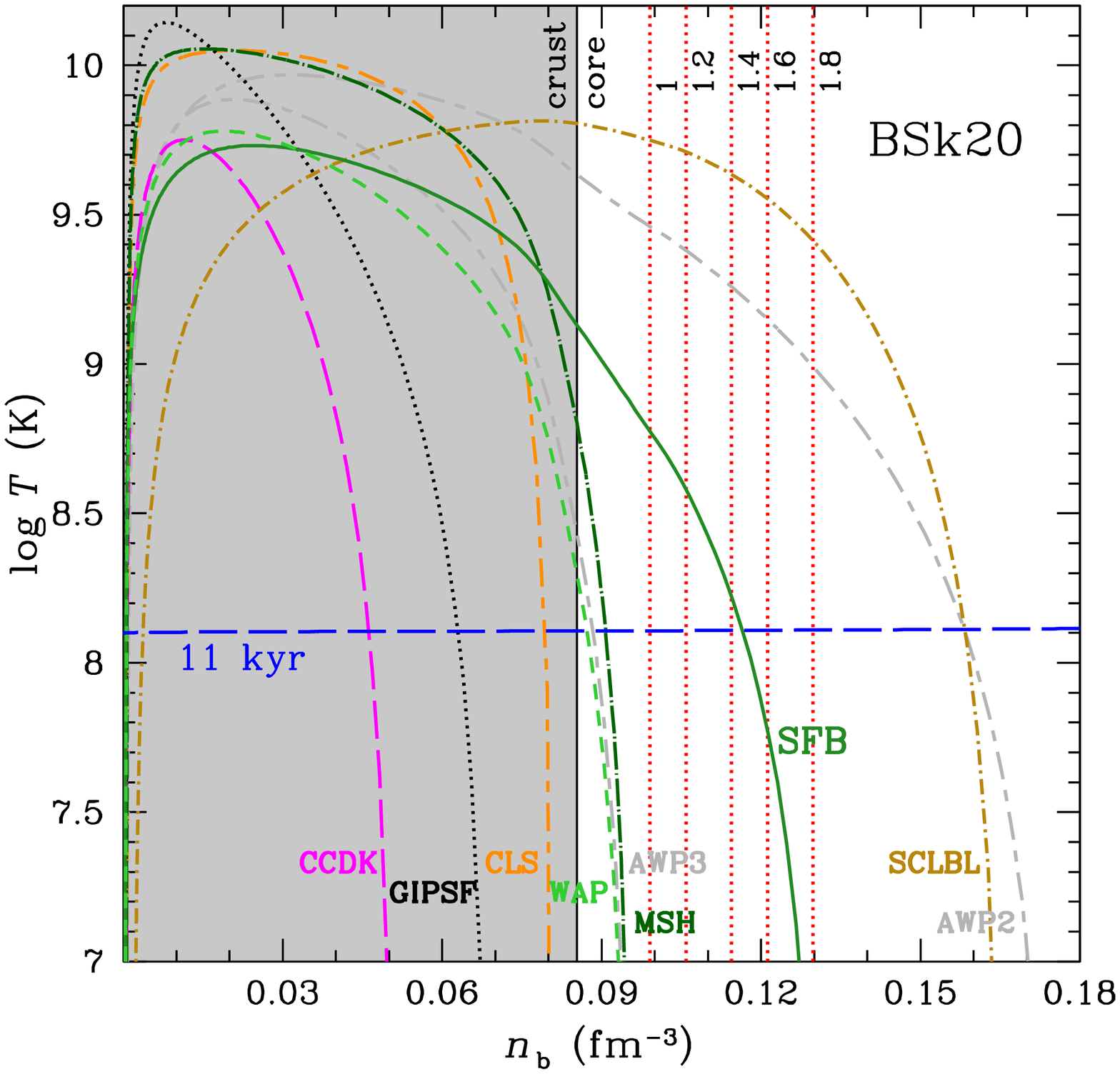}
\hspace{1.5cm}
\includegraphics[scale=0.30]{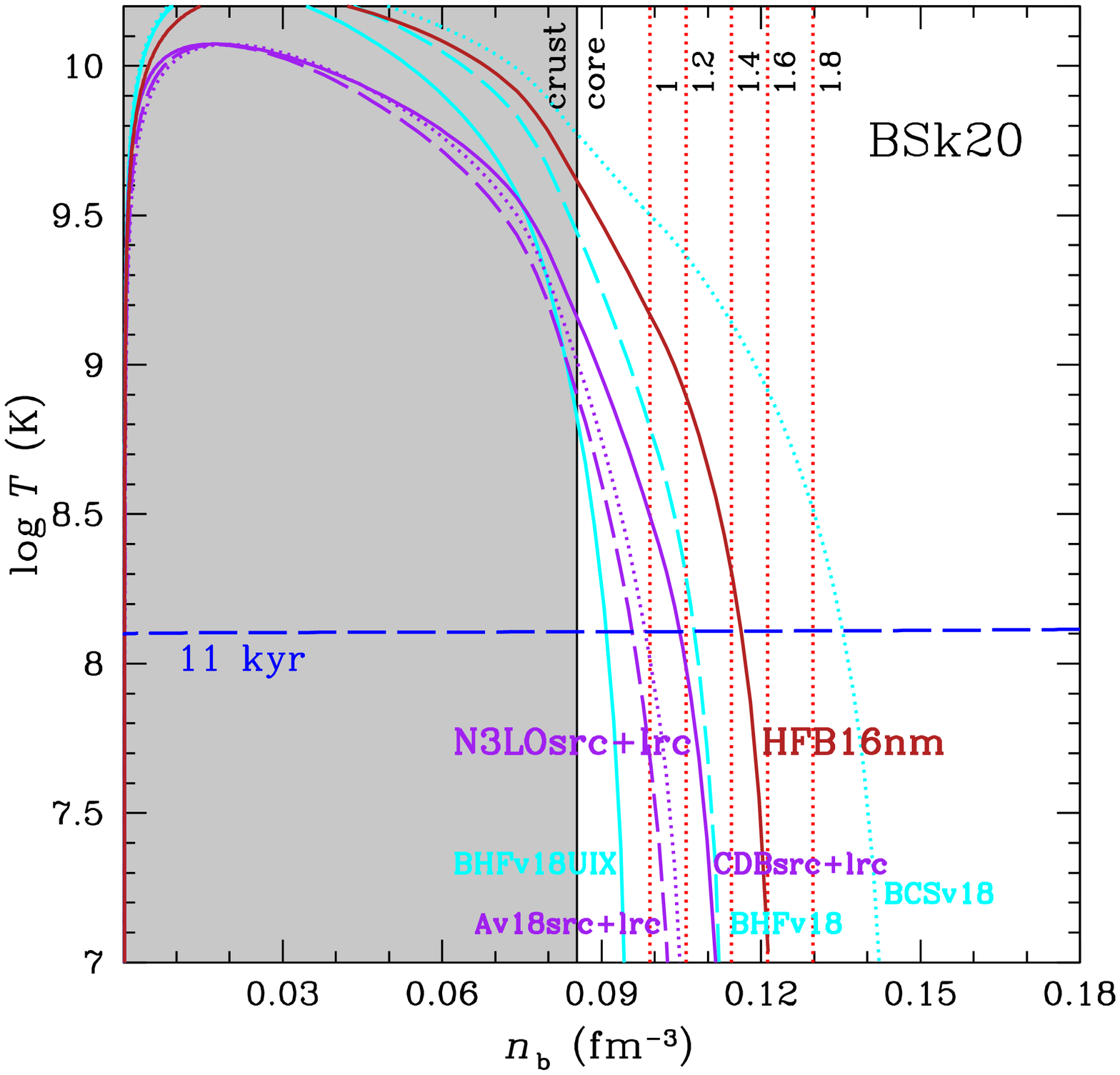}
\caption{
Temperature as a function of baryon number density, using the BSk20 EOS.
Curved lines are the superfluid critical temperature for nine models
from \cite{hoetal15} (left panel) and seven of eleven models discussed in
Sec.~\ref{sec:new} (right panel).
The vertical solid line indicates the separation between the crust
(shaded region) and the core.
Vertical dotted lines denote the density at which the partial moment of
inertia is Vela's $G=1.6\%$ of the total stellar moment of inertia for
neutron stars of different mass (labeled in units of solar mass).
The (nearly horizontal) dashed line is the temperature of a $1.4\,M_{\rm Sun}$
neutron star at the Vela pulsar's age of 11000 years.}
\label{fig:tempdens}
\end{center}
\vspace*{-1.0cm}
\end{figure}

In our recent work \cite{hoetal15b}, we examine a large new dataset of
glitches \cite{espinozaetal11,yuetal13} and explore an idea proposed in
\cite{anderssonetal12} and illustrated in Fig.~\ref{fig:tempdens}.
The neutron star crust (core) is denoted by shaded (unshaded) regions.
Vertical dotted lines indicate the density at which the moment of inertia
exceeds $G = 1.6\%$ (for the Vela pulsar) for neutron star models of various
mass.  Low mass neutron stars have a thicker crust than high mass stars, and
we see that a neutron superfluid confined to the crust can only provide
a relative moment of inertia of 1.6\% for neutron stars with a mass much
less than $1\,M_{\rm Sun}$.
Furthermore, all glitching pulsars with $G\approx 1\%$ would have to be of
low mass ($<1\,M_{\rm Sun}$) as well.
For typical neutron star masses of $1.2$ to $2\,M_{\rm Sun}$ \cite{lattimer16},
some small fraction of the core must contribute to the moment of inertia
required by glitches seen in, for example, the Vela pulsar.

\subsection{Question 2: How much superfluid moment of inertia do pulsars
 possess?}

In the previous analyses of
\cite{linketal99,anderssonetal12,chamel13,piekarewiczetal14,steineretal15,delsateetal16,lietal16},
pulsar glitches are assumed to tap the angular momentum reservoir
associated with superfluid neutrons in the inner crust of the star
(and largely ignoring the temperature dependence of superfluidity).
Therefore, by calculating the entire moment of inertia of the crust,
it is possible to determine the maximum reservoir available for producing
glitches, and this is found to be smaller than that needed to explain
observed glitch activity \cite{anderssonetal12,chamel13,delsateetal16,lietal16}
due to the effect of entrainment \cite{chamel12},
unless the crust is unusually thick \cite{piekarewiczetal14,steineretal15}.
Here and in \cite{hoetal15b}, we consider a superfluid reservoir that extends
into the stellar core
and account for the temperature dependence of superfluidity.
Importantly, we use the observed temperature of pulsars to constrain the latter.
Figure~\ref{fig:tempdens} shows models of critical temperature of neutron
(singlet-state) superfluidity as a function of baryon number density
$n_{\rm b}$:
models plotted in the left panel are from \cite{hoetal15},
while those in the right panel are discussed in Sec.~\ref{sec:new}.
For some superfluid models, the critical temperature, and hence allowed region
for neutrons to become superfluid, is confined to the inner crust,
i.e., in the shaded region to the left of the vertical solid line.
Therefore, pulsar glitches can only involve the moment of inertia
of the inner crust if one of these superfluid models is the correct one.
However, there are superfluid models that extend into the core,
e.g., solid curve labeled SFB or HFB-16nm.
For superfluid models such as the SFB model, if the pulsar temperature is
low enough so that neutrons in the inner crust {\it and} outer core are
superfluid, then pulsar glitches could involve additional moment of inertia
from the core.

\subsection{Question 3: How much superfluid moment of inertia do observed
 pulsars have at present time?}

Finally, are enough neutrons in the crust and core actually superfluid, i.e.,
to the left of one of the vertical dotted lines {\it and} below the superfluid
critical temperature $T_{\rm c}$ in Fig.~\ref{fig:tempdens}?
To answer this question, we need to determine the interior temperature
$T(n_{\rm b})$ of a neutron star and evaluate at what
densities $n_{\rm b}$ the inequality $T<T_{\rm c}$ is satisfied.
This will vary for each
pulsar, depending on its age and/or measured temperature.
Neutron stars are born in supernovae at very high temperatures but
cool rapidly because of efficient neutrino emission.
We perform neutron star cooling simulations using standard
neutrino emission processes to find the interior temperature at various ages
\cite{hoetal12}.
The dashed lines in Fig.~\ref{fig:tempdens} show the resulting temperature
profile (at the age of the Vela pulsar) for a $1.4\,M_{\rm Sun}$ neutron star.
The temperature profile is different for different mass but not
drastically so, unless neutrino emission by direct Urca
processes occurs.
We find that, among nine superfluid models which span a wide range in
parameter space (see left panel of Fig.~\ref{fig:tempdens}),
only the SFB model provides a superfluid reservoir of the required level
(see below for more recent results).
For superfluid models that are confined to the crust, the reservoir is too
small, whereas the reservoir is too large for models that extend much deeper
into the core.  The latter would be unable to explain the regularity of
similar-sized glitches, which requires the reservoir to be completely
exhausted in each event.

\section{Measuring Pulsar Masses and New Results Since Ho et al.~2015
 \cite{hoetal15b}} \label{sec:new}

\begin{figure}[tbh]
% \vspace*{-0.6cm}
\begin{center}
\includegraphics[scale=0.30]{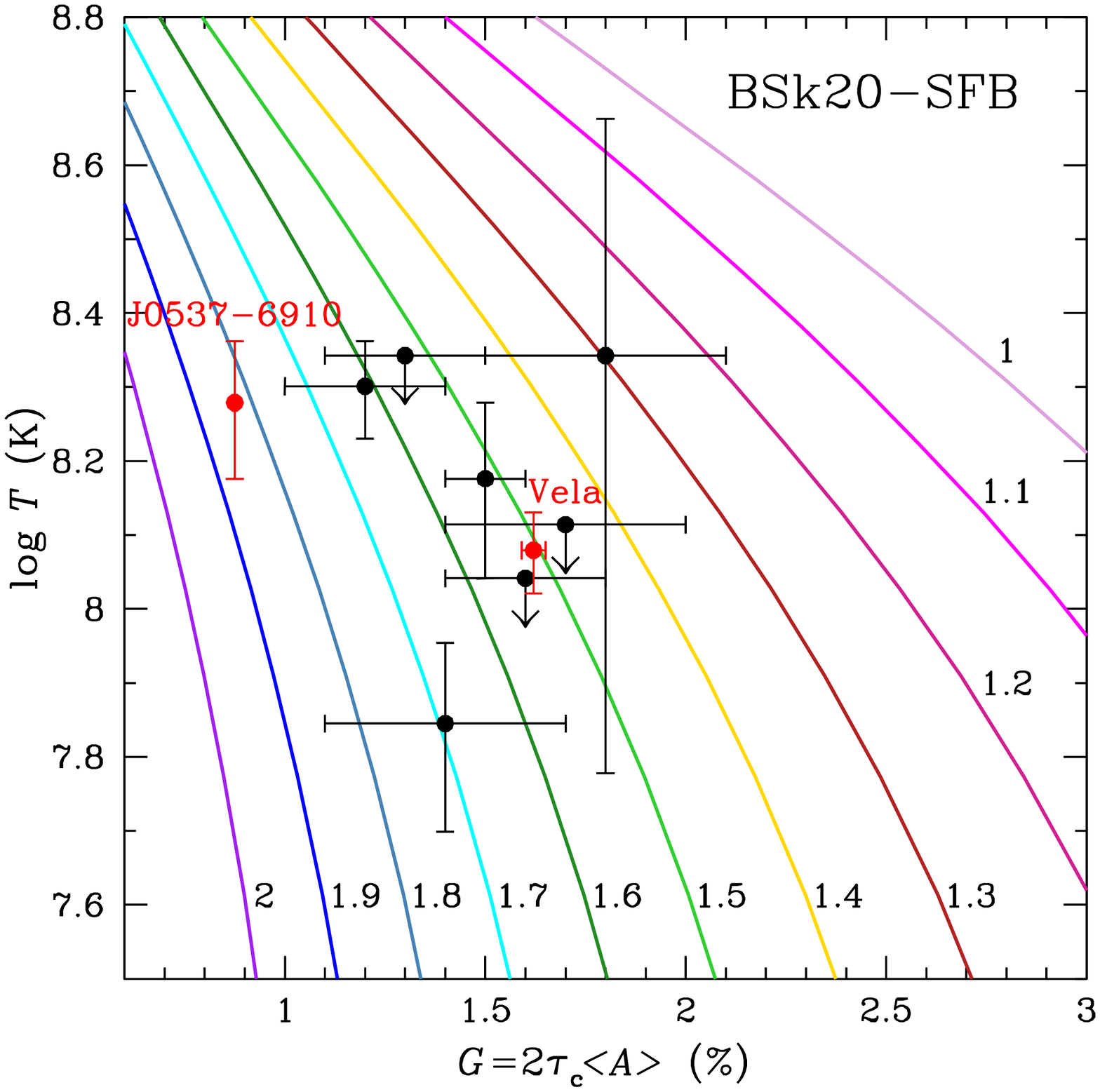}
\hspace{1.5cm}
\includegraphics[scale=0.30]{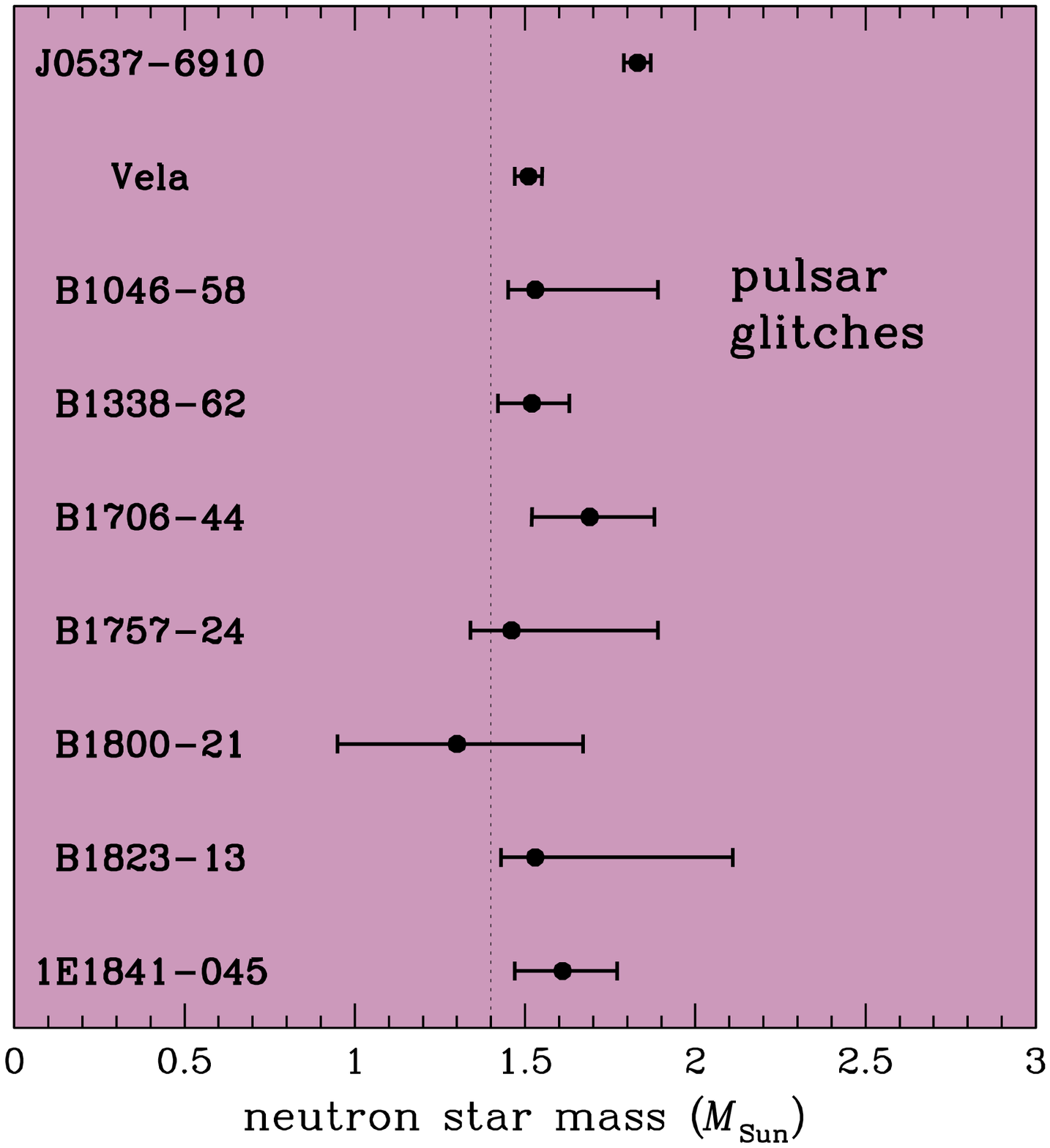}
\caption{
Left panel: Neutron star mass from pulsar observables $G$ and interior
temperature $T$.  Data points are for pulsars with measured $G$ from glitches
and $T$ from an age or surface temperature observation \cite{hoetal15b}.
Lines (labeled by neutron star mass, in units of solar mass) are the
theoretical prediction for $G$ and $T$ using the BSk20 EOS and SFB
superfluid models.
Right panel: Neutron star mass and 1$\sigma$ uncertainty
using the BSk20 EOS and SFB superfluid models;
see Table~2 in \cite{hoetal15b} for masses derived using APR and BSk21
EOS models.}
\label{fig:mass}
\end{center}
\vspace*{-0.7cm}
\end{figure}

The intersection of the three lines in Fig.~\ref{fig:tempdens}
[vertical dotted line at $1.4\,M_{\rm Sun}$ for glitch requirement $G=1.6\%$,
solid line for SFB model of superfluid critical temperature $T_{\rm c}$,
and horizontal dashed line for neutron star temperature $T(n_{\rm b})$ at age
$=11000\mbox{ years}$] is one of our key findings:
{\it  The Vela pulsar has a mass near the characteristic value of
$1.4\,M_{\rm Sun}$, and the size and frequency of Vela's observed glitches
are a natural consequence of the superfluid moment of inertia available to
it at its current age}.
Our results using the BSK20 EOS and SFB superfluid models are summarized
in the left panel of Fig.~\ref{fig:mass}, which shows interior temperature $T$
of a pulsar as a function of glitch parameter $G$
(see \cite{hoetal15b} for results using APR and BSk21 EOS models).
Note that $G$ and $T$ are directly determined from observational data.
The former comes from radio or X-ray glitch measurements.
The latter is obtained either from the age of the pulsar or
by measuring the surface temperature of the pulsar through X-ray observations.
The age gives the interior temperature via neutron star cooling simulations,
whereas surface temperature is related to interior temperature via well-known
relationships \cite{hoetal12}.
Thus for a given pulsar that has measured $G$ and $T$, Fig.~\ref{fig:mass}
allows one to determine the pulsar's mass.
For example, using the BSk20 EOS and SFB superfluid models, we find that Vela
is a $1.51\pm0.04\,M_{\rm Sun}$ neutron star and PSR~J0537$-$6910 is a
$1.83\pm0.04\,M_{\rm Sun}$ neutron star.
Results for seven other glitching pulsars are also plotted in
Figure~\ref{fig:mass}.

Since \cite{hoetal15b}, we tested an additional two EOS models and eleven
superfluid models (see right panel of Fig.~\ref{fig:tempdens}):
two EOS models from \cite{baldoetal14}, along with superfluid
models calculated using the same interactions \cite{zhouetal04};
same HFB-16 symmetric and neutron matter superfluid models that are
used to construct BSk EOS models \cite{gorielyetal09};
and six superfluid models from \cite{dingetal16}.
In brief, the models of \cite{zhouetal04,baldoetal14} imply low
masses, while superfluid models HFB-16nm \cite{gorielyetal09} and
N3LOsrc+lrc \cite{dingetal16} yield reasonable masses.

The ability to measure the mass of isolated pulsars has not been previously
demonstrated.  The most precise neutron star mass measurements to date are
by radio timing of pulsars that are in a binary star system \cite{lattimer16}.
Our method of using glitches to measure mass can greatly increase the
number of known masses, thereby providing constraints on fundamental
physics properties such as the nuclear EOS and superfluidity.
Although there are currently relatively large systematic uncertainties,
these will improve as the understanding of dense matter improves.
The novelty of our approach is the combination of pulsar glitch data and
the temperature dependence of superfluidity.
The method is especially promising with upcoming astronomical observatories
such as the Square Kilometer Array (SKA) in radio and Athena+ in X-rays.
SKA could discover all observable pulsars in the Galaxy, and a program to
monitor glitching pulsars could transform the fields of neutron star and
nuclear physics.

\vspace{0.2cm}
WCGH thanks M. Baldo, F. Burgio, N. Chamel, K. Glampedakis, A. Rios,
and H.-J. Schulze for providing models and for discussion.
WCGH acknowledges support from UK STFC.


\vspace{-0.2cm}
\begin{thebibliography}{99}
\bibitem{andersonitoh75}
 P.~W. Anderson, N. Itoh,
 Nature {\bf 256}, 25 (1975).
\bibitem{baymetal69}
 G. Baym, C. Pethick, D. Pines, M. Ruderman,
 Nature {\bf 224}, 872 (1969).
\bibitem{hoetal15b}
 W.~C.~G. Ho, C.~M. Espinoza, D. Antonopoulou, N. Andersson,
 Science Advances {\bf 1}, e1500578 (2015).
\bibitem{espinozaetal11}
 C.~M. Espinoza, A.~G. Lyne, B.~W. Stappers, M. Kramer,
 Mon. Not. R. Astron. Soc. {\bf 414}, 1679 (2011).
\bibitem{yuetal13}
 M. Yu, et al.,
 Mon. Not. R. Astron. Soc. {\bf 429}, 688 (2013).
\bibitem{anderssonetal12}
 N. Andersson, K. Glampedakis, W.~C.~G. Ho, C.~M. Espinoza,
 Phys. Rev. Lett. {\bf 109}, 241103 (2012).
\bibitem{lattimer16}
 J.~M. Lattimer,
 Jpn. Phys. Soc. Conf. Proc., this volume.
\bibitem{linketal99}
 B. Link, R.~I. Epstein, J.~M. Lattimer,
 Phys. Rev. Lett. {\bf 83}, 3362 (1999).
\bibitem{chamel13}
 N. Chamel,
 Phys. Rev. Lett. {\bf 110}, 011101 (2013).
\bibitem{piekarewiczetal14}
 J. Piekarewicz, F.~J. Fattoyev, C.~J. Horowitz,
 Phys. Rev. C {\bf 90}, 015803 (2014).
\bibitem{steineretal15}
 A.~W. Steiner, S. Gandolfi, F.~J. Fattoyev, W.~G. Newton,
 Phys. Rev. C {\bf 91}, 015804 (2015).
\bibitem{delsateetal16}
%  T. Delsate, N. Chamel, N. G\"{u}rlebeck, A.~F. Fantina, J.~M. Pearson,
%  C. Ducoin,
 T. Delsate et al.,
 Phys. Rev. D {\bf 94}, 023008 (2016).
\bibitem{lietal16}
 A.~Li, J.~M. Dong, J.~B. Wang, R.~X. Xu,
 Astrophys. J. Suppl. {\bf 223}, 16 (2016).
\bibitem{chamel12}
 N. Chamel,
 Phys. Rev. C {\bf 85}, 035801 (2012).
\bibitem{hoetal15}
 W.~C.~G. Ho, K.~G. Elshamouty, C.~O. Heinke, A.~Y. Potekhin,
 Phys. Rev. C {\bf 91}, 015806 (2015).
\bibitem{hoetal12}
 W.~C.~G. Ho, K. Glampedakis, N. Andersson,
 Mon. Not. R. Astron. Soc. {\bf 422}, 2632 (2012).
\bibitem{baldoetal14}
 M. Baldo, G.~F. Burgio, H.-J. Schulze, G. Taranto,
 Phys. Rev. C {\bf 89}, 048801 (2014).
\bibitem{zhouetal04}
 X.-R. Zhou, H.-J. Schulze, E.-G. Zhao, F. Pan, J.~P. Draayer,
 Phys. Rev. C {\bf 70}, 048802 (2004).
\bibitem{gorielyetal09}
 S. Goriely, N. Chamel, and J.~M. Pearson,
 Eur. Phys. J. A {\bf 42}, 547 (2009).
\bibitem{dingetal16}
% D. Ding, A. Rios, H. Dussan, W.~H. Dickhoff, S.~J. Witte, A. Carbone, A. Polls,
 D. Ding et al.,
 Phys. Rev. C {\bf 94}, 025802 (2016).
\end{thebibliography}
\end{document}